\documentstyle[aps,twocolumn,amssymb]{revtex}
\def\Diff{{\rm Diff}}
\def\beq{\begin{equation}}
\def\eeq{\end{equation}}
\def\bea{\begin{eqnarray}}
\def\eea{\end{eqnarray}}
\def\ba{\begin{array}}
\def\ea{\end{array}}
\def\pa{\partial}

\def\asr{a_{\sigma R}}
\def\asrd{a_{\sigma R}^\dagger}
\def\asl{a_{\sigma L}}
\def\asld{a_{\sigma L}^\dagger}
\def\ga{\gamma}

\def\si{\sigma}

\def\de{\delta}

\def\Om{\Omega}

\def\rf{{\rm f}}\def\la{\lambda}

\def\eps{\epsilon}
\def\rd{{\rm d}}
\def\ri{{\rm i}}
\def\sgn{{\rm sgn}}

\def\hsi{{\hat \si}}
\def\hgR{{\hat g_R}}
\def\hgL{{\hat g_L}}
\def\hgaL{{\hat g}^\ast_L}
\def\hgaR{{\hat g}^\ast_R}

\def\hg0{{\hat g(0)}}
\def\hga0{{\hat g}^\ast(0)}
\def\ln{{\rm ln}}
\def\sgn{{\rm sgn}}

\def\nn{\nonumber}
\def\e{{\rm e}}
\def\cH{{\mathcal H}}
\def\cF{{\mathcal F}}
\def\Weyl{{\rm Weyl}}
\def\Aut{{\rm Aut}}

\renewcommand\Im{{\rm Im}}
\begin{document}
\draft
\twocolumn\twocolumn[\hsize\textwidth\columnwidth\hsize\csname
@twocolumnfalse\endcsname
\title{Quantum massless field in 1+1 dimensions}
\author{Jan Derezi{\'n}ski${}^a$, Krzysztof A. Meissner${}^b$}
\address{
${}^a$ Dept of Math. Methods in Physics,
Warsaw University,\\
Ho\.za 74,
00-682 Warsaw, Poland,
\\
${}^b$ Institute of Theoretical Physics,
 Warsaw University\\ Ho\.za 69,
00-681 Warsaw, Poland}
\maketitle

\begin{abstract}
We present a construction of the algebra of operators and the Hilbert
space for a quantum massless field in 1+1 dimensions.
\end{abstract}
\pacs{} ]

\section{Introduction}

It is usually stated that quantum massless bosonic fields in 1+1
dimensions (with noncompact space dimension) do not exist. With
massive fields the correlation function
\beq
\langle\Om|\phi(f_1)\phi(f_2)\Om\rangle=\int\frac{\rd
p}{2\pi 2E_p}\hat f_1^\ast(E_p,p)\hat f_2(E_p,p)
\eeq
(where $E_p=\sqrt{p^2+m^2}$) is well defined but in the limit $m\to
0$ diverges because of the infrared problem.
The limit exists only after adding an additional nonlocal
constraint on the smearing functions:
\beq
\hat f(0,0)=\int\rd t\rd x f(t,x)=0.
\label{cons}
\eeq
Under this constraint it is not difficult to construct massless
fields in 1+1 dimension (see eg. \cite{StW}, where the framework
of the Haag-Kastler axioms is used).

Massless fields are extensively used for example in string theory
(albeit most often after Wick rotation to the space with Euclidean
signature). They also appear as the scaling limit of massive
fields \cite{BV}. Usually, in these applications, the constraint
(\ref{cons}) appears to be  present at least implicitly. e.g. in
string amplitudes one imposes the condition that sum of all
momenta is equal to 0. Nevertheless, it seems desirable to have a
formalism for massless 1+1-dimensional fields free of this
constraint.

In the literature there are many papers that propose to use an
indefinite metric Hilbert space for this purpose
\cite{IZ,Na1,Na2,GKW,MPS,DBR}. Clearly, an indefinite metric is
not physical and in order to determine physical observables one
needs to perform a reduction similar to that of the Gupta-Bleuler
formalism used in QED. The outcome of this Gupta-Bleuler-like
procedure is essentially equivalent to imposing the constraint
(\ref{cons}) \cite{MPS}. Therefore, we do not find the indefinite
metric approach appropriate.

In this paper we present two explicit constructions of  (positive
definite) Hilbert spaces with  representations of the massless
Poincar\'e algebra in 1+1 dimensions and local  fields (or at
least their exponentials). We allow
all  test functions $f$ that belong to the Schwartz class on the 1+1
dimensional Minkowski space,   without the constraint (\ref{cons}).
We try to make sure that as many Wightman axioms as possible are satisfied.

In the first construction we obtain a separable Hilbert space and well defined
fields, however we do
not have a vacuum vector. In the second construction, the Hilbert space is
non-separable, only  exponentiated fields  are well defined, but there
exists a vacuum vector. Thus, neither of them satisfies
 all Wightman axioms. Nevertheless,
we believe that both our constructions  are good  candidates for
a physically correct massless quantum
field theory in 1+1 dimensions.

Our constructions have supersymmetric extensions, which we
describe at the end of our article.

In the literature known to us the only place where one can find a
treatment of massless fields in  1+1 dimension similar to ours is
\cite{AMS1,AMS2} by Acerbi, Morchio and Strocchi.
Their construction is equivalent to our second (nonseparable)
construction. We have never seen  our first (separable)
construction of massless fields in the literature.

Acerbi, Morchio and Strocchi start from the $C^*$-algebra
associated to the CCR over the symplectic space of solutions of
the wave equation parametrized by the initial conditions. Then
they apply the GNS construction to the Poincar\'{e} invariant
quasi-free state obtaining a non-regular representation of CCR.

In our presentation we prefer to use the derivatives of right and
left movers to parametrize fields, rather than the initial
conditions. We also avoid, as long as possible, to invoke abstract
constructions from the theory of $C^*$-algebras, which may be less
transparent to some of the  readers. We explain the relationship
between our formalism and that of \cite{AMS1,AMS2}. The symmetry 
structure of this theory is surprisingly rich. Some of the 
objects are covariant only under Poincar\'e group but there are others 
that are covariant under larger groups:  
$A_+(1,{\mathbb R})\times A_+(1,{\mathbb R})$,
$SL(2,{\mathbb R})\times SL(2,{\mathbb R})$, $\Diff_+({\mathbb R})\times 
\Diff_+({\mathbb R})$,  $\Diff_+(S^1)\times  
\Diff_+(S^1)$.

\section{Fields}

The action of the 1+1 dimensional free real scalar massless field
theory reads
\beq
S= \frac12 \int \rd t\rd x \ \left(
(\pa_{t}\phi)^2-(\pa_x\phi)^2\right) .
\label{action}
\eeq
This leads to the  equations of motion
\beq
 (-\pa_t^2+\pa_x^2)\phi=0.
\label{eqofm}
\eeq
The solution of (\ref{eqofm}) is the sum of right and left movers,
i.e. functions of $(t-x)$ and $(t+x)$ respectively:
\beq
 \phi(t,x)=\phi_R(t-x)+\phi_L(t+x).
\label{phisol}
\eeq


We will often used ''smeared'' fields in the sense
\[
\phi(f)=\int \rd t \rd x\, \phi(t,x)f(t,x),
\]
where  we assume that $f$ are real Schwartz functions.
Because of (\ref{phisol}), they can be written in the form
\[
\phi(f)=\phi( g_R, g_L),
\]
where
\[
\hgR(k)=\hat f(k,k),\ \ \ \ \hgL(k)=\hat f(k,-k),
\]
($k\ge0$) and the Fourier transforms of the test function $f$ and $g$
are defined as
\beq
\hat f(E,p):=\int\rd t\rd x \ f(t,x)\ \e^{\ri Et-\ri px},
\eeq
\beq
\hat g(k):=\int_{-\infty}^{\infty}\rd t g(t)\e^{-\ri kt}.
\eeq
The function $g_R$ corresponds to right movers and $g_L$ to left
movers. Note that
\beq
\hgR(0)=\hgL(0)=:\hat g(0).
\label{zero}
\eeq
$\hg0$ is real, because function $f$ is real.

We introduce the notation
\bea
&&( g_1| g_2):=\label{scprod}\\&&
\ \ \ \frac{1}{2\pi}\lim_{\eps\searrow
0}\left(\int\limits_{\eps}^\infty\frac{\rd k}{k}
 {\hat g}^\ast_1(k) {\hat g}_2(k)
 +\ln(\eps/\mu) {\hat g}^\ast_1(0) {\hat g}_2(0)\right),\nn
\eea
where $\mu$ is a positive constant having the dimension of mass.
For functions that satisfy $\hg0=0$, $( g_1| g_2)$ is a
(positive) scalar product -- otherwise it is not positive definite
and therefore cannot be used directly in the construction of a
Hilbert space. Such a scalar product corresponds to quantization
of the theory in a constant compensating background.

In view of the infrared divergence we factorize the Hilbert space
into two parts -- one that is infrared safe and the second that in
some sense regularizes the divergent part.

We introduce the creation $a^\dagger_R(k),\ a^\dagger_L(k)$ and
annihilation $a_R(k),\ a_L(k)$ operators as well as pairs of
operators $(\chi,p^\dagger)$, $(\chi^\dagger,p)$. They satisfy the
commutation relations
\bea
\left[a_R(k),a^\dagger_R(k')\right]&=&2\pi k\de(k-k'),\nn\\
\left[a_L(k),a^\dagger_L(k')\right]&=&2\pi k\de(k-k'),\nn\\
\left[\chi,p\right]&=&\ri
\label{alg}
\eea
with all other commutators vanishing.

To proceed we choose two real functions $\si_R(x)$ and $\si_L(x)$
satisfying
\beq
 \hsi_R(0)=\hsi_L(0)=1
\label{sidef}
\eeq
and otherwise arbitrary. To simplify further formulae we define the
combinations
\bea
\asr(k)&:=& a_R(k)-\ri\hsi_R(k)\chi\nn\\
\asrd(k)&:=& a^\dagger_R(k)+\ri\hsi^\ast_R(k)\chi\nn\\
\asl(k)&:=& a_L(k)-\ri\hsi_L(k)\chi\nn\\
\asld(k)&:=& a^\dagger_L(k)+\ri\hsi^\ast_L(k)\chi
\label{defb}
\eea
and therefore
\bea
\left[\asr(k),\asrd(k')\right]&=&2\pi k\de(k-k'),\nn\\
\left[\asl(k),\asld(k')\right]&=&2\pi k\de(k-k'),\nn\\
\left[\asr(k),p\right]&=&\hsi_R(k),\nn\\
\left[\asrd(k),p\right]&=&-\hsi^\ast_R(k),\nn\\
\left[\asl(k),p\right]&=&\hsi_L(k),\nn\\
\left[\asld(k),p\right]&=&-\hsi^\ast_L(k).
\eea

Now we are in a position to introduce the field operator $\phi(
g_R, g_L)$, depending on a pair of functions $g_R$, $g_L$
satisfying (\ref{zero}).

\bea
 \phi(g_R, g_L)
&=&\int\frac{\rd k}{2\pi k}
 \Big((\hgR(k)-\hg0\hsi_R(k))\asrd(k)\nn\\
 && +(\hgaR(k)-\hg0\hsi^\ast_R(k))\asr(k)
 \nn\\
&& +(\hgL(k)-\hg0\hsi_L(k))\asld(k)\nn\\
&&+(\hgaL(k)-\hg0\hsi^\ast_L(k))\asl(k)\Big)\nn\\
&&+\hg0 p.
\label{phigen}
\eea

The field $\phi(g_R,g_L)$ is  hermitian and satisfies the commutation
relation
\bea&&
 \left[\phi( g_{R1}, g_{L1}),\phi( g_{R2}, g_{L2})\right]\nn\\
&=&(g_{R1}|g_{R2})
-(g_{R2}|g_{R1})\nn\\
&&+(g_{L1}|g_{L2})-(g_{L2}|g_{L1})\nn\\
&=&\ri 2\Im(g_{R1}|g_{R2})+\ri 2\Im(g_{L1}|g_{L2}).
\label{phig}
\eea
The commutator in (\ref{phig}) does not depend on the
functions $\sigma_R$, $\sigma_L$.

\section{Poincar\'{e} covariance}

Let $A_+(1,{\mathbb R})$ denote the group of orientation preserving affine 
transformations of the real
line, that is the group of maps $t\mapsto at+b$ with $a>0$. 
The proper Poincar\'e group in 1+1 dimension can be naturally embedded in
the direct product of two copies of  $A_+(1,{\mathbb R})$, one for
the right movers and one for the left movers.

The infinitesimal
generators of the right $A_+(1,{\mathbb R})$ group will be denoted $H_R$
(the right Hamiltonian) and $D_R$ (the right generator of
dilations) and they satisfy the commutation relations
\[[D_R,H_R]=\ri H_R.\]

The representation of these operators in terms of the creation and
annihilation operators is given by
\bea
 H_R&=&\int \frac{\rd k}{2\pi}\  \asrd(k) \asr(k),\nn\\
 D_R&=&\frac{\ri}{2} \int \frac{\rd k}{2\pi}\left(\asrd(k) \pa_k \asr(k)-
\bigl(\pa_k \asrd(k)\bigr)  \asr(k)\right),\nn\\
\eea
Their action on fields is given by
\bea
 \left[H_R,\phi( g_R, g_L)\right]&=&-\ri\phi(\pa_tg_R,0),\nn\\
  \left[D_R,\phi(g_R,g_L)\right]&=&\ri\phi(\pa_tt g_R,0),
 \nn\eea
and in the exponentiated form by
\bea
\e^{\ri sH_R}\phi(g_R,g_L)\e^{-\ri sH_R}&=&\phi(g_R(\cdot-s),g_L,
\nn
\\
\e^{\ri s D_R}\phi(g_R,g_L)\e^{-\ri s D_R}&=&\phi(\e^{-s}g_R(\e^{-s}\cdot
),g_L ).\nn
\eea

For $(a,b)\in A_+(1,{\mathbb R})$ 
we set $r_{a,b}g(t):=a^{-1}g(a^{-1}(t-b))$ and
\[R_R(a,b)=\e^{\ri\ln a D_R}\e^{\ri b H_R}.\]
$R_R$ is a unitary representation of
$ A_+(1,{\mathbb R})$, which acts naturally on the fields:
\bea &&R_R(a,b)\phi(g_R,g_L)R_R(a,b)^\dagger\nn\\
&=&\phi(r_{a,b}g_R,g_L).\label{koko}\eea
Note, however, that $r_{a,b}$ does not preserve the indefinite scalar product
(\ref{scprod}) unless we impose the constraint $\hat g(0){=}0$:
\bea
(r_{a,b}g_1|r_{a,b}g_2)&=&
(g_1|g_2)-\ln a
\hat g_1^*(0)\hat g_2(0).\nn\eea

Similarly we introduce the left Hamiltonian $H_L$ and the left
generator of dilations $D_L$ satisfying analogous commutation relations
 and the
representation of the left $ A_+(1,{\mathbb R})$.

The Poincar\'e{} group  generators are the Hamiltonian
$H=H_R+H_L$, the momentum $P=H_R-H_L$ and the boost operator
$\Lambda=D_R-D_L$ (the only Lorentz generator in 1+1 dimensions).
The elements of the Poincar\'{e} group are  of the form
\[(a,b_R),(a^{-1},b_L)\in A_+(1,{\mathbb R})\times A_+(1,{\mathbb R}).\]
The scalar product $(g_{R1}|g_{R2})+(g_{L1}|g_{L2})$ is invariant wrt the
proper Poincar\'e{}  group.

\section{Changing the compensating functions}

It is important to discuss the dependence of the whole
construction on the choice of compensating functions $\si_R$ and
$\si_L$.

Let $\tilde\si_R$ and $\tilde\si_L$ be another pair of real functions  
satisfying (\ref{sidef}).
Set $\xi_R(x):=\tilde\si_R(x)-\si_R(x)$, $\xi_L(x):=\tilde\si_L(x)-\si_L(x)$.
Note that $\hat\xi_R(0)=\hat\xi_L(0)=0$. Define
\bea
U(\xi_R,\xi_L)&=&\exp\left(\int\frac{\rd k}{2\pi k}\right.\nn\\
&&\!\!\!\Bigl(\ri\chi\hat\xi^\ast_R(k)a_R(k)+
\ri\chi\hat\xi_R(k)a^\dagger_R(k)\nn\\
&&\!\!\!+\frac12\chi^2(\hat\xi^\ast_R(k)\hsi_R(k)
-\hat\xi_R(k)\hsi^\ast_R(k))\nn\\
&&\!\!\!+ R\rightarrow L\Bigr)\Big).
\label{gauge}
\eea
 Using the formula
\beq
\e^A B\e^{-A}=B+[A,B]+\frac12[A,[A,B]]+\ldots
\eeq
we have
\bea
Ua_R(k)U^{-1}&=&a_R(k)-\ri\chi\hat\xi_R(k),\nn\\
Ua^\dagger_R(k)U^{-1}&=&a^\dagger_R(k)+\ri\chi\hat\xi^\ast_R(k),\nn\\
Ua_L(k)U^{-1}&=&a_L(k)-\ri\chi\hat\xi_L(k),\nn\\
Ua^\dagger_L(k)U^{-1}&=&a^\dagger_L(k)+\ri\chi\hat\xi^\ast_L(k),\nn\\
U\chi U^{-1}&=&\chi
\eea
and
\bea
UpU^{-1}&=&p+\int\frac{\rd k}{2\pi k}\Big(-\hat\xi^\ast_R(k)a_R(k)
-\hat\xi_R(k)a_R^\dagger(k)\nn\\
&&-\ri\chi\hat\xi_R(k)\hsi^\ast_R(k)+
\ri\chi\hat\xi^\ast_R(k)\hsi_R(k)\nn\\
&&\left.+
R\rightarrow L\right).
\eea
Using these relations we get for example
\bea
U\phi_\sigma(g_R,g_L)U^{-1}&=&\phi_{\tilde\sigma}(g_R,g_L),\nn\\
U \asrd U^{-1}&=& a_{\tilde\sigma R}^\dagger,\nn\\
U H_{\sigma R} U^{-1}&=& H_{\tilde  \sigma R},
\label{upoleu}
\eea
where we made explicit the dependence of $\phi$ and $H_R$
on $\sigma$ and $\tilde\sigma$. Thus the two constructions
 --- with $\sigma$ and with $\tilde\sigma$ --- are unitarily equivalent.

\section{Hilbert space}
\label{sec.hil}

The Hilbert space of the system is the product of three spaces:
${\mathcal H}={\mathcal H}_R\otimes{\mathcal H}_L\otimes{\mathcal
H}_0$. ${\mathcal H}_R$ is the bosonic Fock space spanned by the
creation operators $a^\dagger_R(k)$ acting on the vacuum vector
$|\Omega_R\rangle$. Analogously ${\mathcal H}_L$ is the bosonic Fock
space spanned by the creation operators $a^\dagger_L(k)$ acting on
the vacuum vector $|\Omega_L\rangle$. With the third sector ${\mathcal
H}_0$ we have essentially two options. If we take the usual choice ${\mathcal
H}_0=L^2({\mathbb R}, \rd\chi)$ then we can define the vacuum state
(vacuum expectation value) but there does not exist a
vacuum vector. On the other hand, we can take ${\mathcal
H}_0=l^2({\mathbb R})$, i.e. the space with the
scalar product
\beq
(f|g)=\sum_{\chi\in{\mathbb R}} f^\ast(\chi)g(\chi),
\eeq
which is a nonseparable space. It may sound as a nonstandard
choice, it has however the advantage of possessing a vacuum
vector. The orthonormal basis in the latter space consists of the
Kronecker delta functions $\de_{\chi}$ for each $\chi\in{\mathbb
R}$. In the nonseparable case, the operator $p$, and
therefore also $\phi(g_R,g_L)$, cannot be defined. But there
exist operators $\e^{ \ri s p}$, for $s\in{\mathbb
R}$, and also $\e^{\ri \phi(g_R,g_L)}$. The commutation relations
for these exponential operators follow from the commutation
relations for $p$ and $\phi(g_R,g_L)$ described
above.

In such a space the vacuum vector is given by
\beq
|\Om\rangle=|\Om_R\otimes\Om_L\otimes\de_0\rangle\,.
\eeq
This vector is invariant under the action of the Poincar\'e group
and the action of the gauge group $U$. We now prove that it is the
unique vector with the lowest energy. Note first that $H$ is
diagonal in $\chi\in{\mathbb R}$. Now for an arbitrary
$\Phi\in{\mathcal H}_R\otimes{\mathcal H}_L$ and $\chi_1\in
{\mathbb R}$,
\bea
&&\langle
\Phi\otimes\delta_{\chi_1}|H|\Phi\otimes\delta_{\chi_1}\rangle\nn\\
&=&\int\Big(\left\langle \Phi|\bigl(a_R^\dagger(k)+
\ri\chi_1\hat\sigma_R^*(k)\bigr)
\bigl(a_R(k)-\ri\chi_1\hat\sigma_R(k)\bigr)\Phi\right\rangle\nn\\&&
+R\to L\Big)
\label{kak}
\eea
For any $\chi_1$, the expression (\ref{kak}) is nonnegative. If
$\chi_1=0$, it has a unique ground state
$|\Omega_R\otimes\Omega_L\rangle$.

If $\chi_1\neq0$, then (\ref{kak}) has no ground state. In fact, it is
well known that a ground state of a quadratic Hamiltonian is a coherent
state, that is given by a vector of the form
\bea
&& |\beta_R,\beta_L\rangle=\nn\\&& C\exp\left(\int\frac{\rd
k}{2\pi k}\left(
\beta_R(k)a^\dagger_R(k) +\beta_L(k)a^\dagger_L(k)\right)\right)
|\Om_R\otimes\Om_L\rangle
\nn
\eea
and $C$ is the normalizing constant
\bea
C=\exp\left(-\frac12\int\frac{\rd k}{2\pi k}
\left(|\beta_R(k)|^2+|\beta_L(k)|^2\right)\right).
\eea
If we set $\Phi=|\beta_R,\beta_L\rangle$ in (\ref{kak}), then we
obtain
\[
\langle\Phi\otimes\delta_{\chi_1}|H|\Phi\otimes\delta_{\chi_1}\rangle
=\left|\beta_R(k)+{\rm i}\chi_1\sigma_R(k)\right|^2+R\to L\,.
\]
that takes the minimum for
\[
\beta_R(k)=-\ri\chi_1\hat\sigma_R(k),\ \ \ \
  \beta_L(k)=-\ri\chi_1\hat\sigma_L(k).
\]
But for $\chi_1\ne 0$, $|\beta_R,\beta_L\rangle$ is not well
defined as a vector in the Hilbert space. To see this we can note
that the normalizing constant $C$ equals zero, because then
\[
\chi_1^2\int\frac{\rd k}{2\pi k}
\left(|\hat\sigma_R(k)|^2+|\hat\sigma_L(k)|^2\right)=\infty.
\]
(The fact that operators of the form (\ref{kak}) have no ground state is well
known in the literature, see eg. \cite{De}).

In the nonseparable case ${\mathcal H}_0=l^2({\mathbb R})$, the
expectation value
\[\langle\Omega|\cdot|\Omega\rangle=:\omega(\cdot)\]
is a Poincar\'e-invariant state (positive linear functional) on
the algebra of observables. If we take the separable case
${\mathcal H}_0=L^2({\mathbb R},\rd\chi)$, the state $\omega$ can
also be given a meaning, even though the vector $\Omega$ does not
exist (since then $\delta_0$ is not well defined).

Note  that in the nonseparable case the state $\omega$ can act on
an arbitrary bounded operator on ${\mathcal H}$. In the separable
case we have to restrict $\omega$ to a smaller algebra of
operators, say, the algebra (or the $C^*$-algebra) spanned by the
operators of the form $\e^{\ri\phi(g_R,g_L)}$.

The expectation values of the exponentials of the 1+1-dimensional
massless field make sense and can be computed, both in the
separable and nonseparable case:
\bea
&&\omega\left(
\exp\left(\ri\phi( g_{R}, g_{L})
\right)
\right)\nn\\
&=&\exp\left(-\frac12\int\frac{\rd k}{2\pi k}\left(
|\hgR(k)|^2+|\hgL(k)|^2\right)\right).
\label{qwq}
\eea
Note that  the integral in the exponent of (\ref{qwq}) is the
usual integral of a positive function, and not its regularization
as in (\ref{scprod}). Therefore, if $\hg0\neq0$, then this
integral equals $+\infty$ and (\ref{qwq}) equals zero.

The  ``two-point functions'' of massless fields in 1+1 dimension,
even smeared out ones, are not well defined. Formally, they are
introduced as
\beq
\omega\left(\phi( g_{R,1}, g_{L,1})
\phi(g_{R,2}, g_{L,2})\right).
\label{2po}
\eeq
If we use the nonseparable ${\mathcal H}_0=l^2({\mathbb R})$, then
field operators $\phi( g_R, g_L)$ is not well defined  if
$\hg0\neq0$, and thus  (\ref{2po}) is not defined. If we use the
separable Hilbert space ${\mathcal H}_0=L^2({\mathbb R},\rd\chi)
$, then $\phi(g_{R,1},g_{L,1})\phi(g_{R,2},g_{L,2})$ are unbounded
operators and there is no reason why the state $\omega$ could act
on them. Thus (\ref{2po}) a priori does not make sense. In the
usual free quantum field theory, if mass is positive or dimension
more than 2,  the expectation values the exponentials depend on
the smeared fields analytically, and by taking their second
derivative at $(g_R,g_L)=(0,0)$,  one can introduce the 2-point
function. This is not the case for massless field in 1+1
dimension.

The above discussion shows that the problem of the non-positive
definiteness of the two-point function, so extensively discussed
in the literature \cite{DBR,Na1,Na2,GKW,MPS} does not exist in our
formalism.

It should be noted that massless fields in 1+1 dimension do not
satisfy the Wightman axioms \cite{SW}. In the separable case there
is no  vacuum vector in the Hilbert space; in the nonseparable
case there is a vacuum vector, but there are no fields $\phi(f)$,
only the ``Weyl operators'' $\e^{\ri\phi(f)}$.

\section{Fields in  position representation}

So far in our discussion we found it convenient to use the momentum
representation. The position representation is, however, better suited for
many purposes.

Let  ${\cal W}(t)$ denote the Fourier transform of the appropriately
regularized distribution  $\frac{\theta(k)}{2\pi k}$, that is
\bea
{\cal W}(t)
&=&
\frac{1}{2\pi}\lim_{\eps\to 0}\left(\int_{k>\eps}\frac{\rd
k}{k}\e^{\ri kt}+\ln(\eps/\mu)\right)\label{wx}\\
&=&\frac{1}{2\pi}\left(-\ga_E-\ln|\mu
t|+\frac{\ri\pi}{2}\sgn(t)\right)={\cal W}^\ast(-t)
\nn
\eea
where $\gamma_E$ is the Euler's constant.
We can rewrite (\ref{scprod}) as
\beq
( g_1| g_2)=\int_{-\infty}^{\infty}\rd t\rd s
 g^\ast_1(t){\cal W}(t-s) g_2(s)
\eeq

To describe massless field in the position representation we
introduce the operators $\psi_R(t)$
defined as
\[
\psi_R(t)=\int\frac{\rd k}{2\pi k}\bigl(a_R^\dagger(k)\e^{-\ri kt}+
a_R(k)\e^{\ri kt}\bigr),
\]
and similarly for $R\to L$.
Note that it is allowed to smear $\psi_R(t)$ and
$\psi_L(t)$ only with test functions satisfying
\[\int g(t)\rd t=0.\]

Note that ${\mathcal W}(t-s)$ and $\frac{\ri}{2}\sgn(t-s)$ are the
correlator and the commutator functions for $\psi_R(t)$: 
\bea
\langle
\Omega_R|\psi_R(t)\psi_R(s)\Omega_R\rangle&=& {\mathcal W}(t-s),\nn
\\[2mm] [\psi_R(t),\psi_R(s)]&=&\frac{\ri}{2}\sgn(t-s),\nn
\eea 
and similarly for $R\to L$.

We introduce also
\bea
\psi_{\sigma R}(t)&
=&\int\frac{\rd k}{2\pi k}\asr^\dagger(k)\e^{-ikt}
+\int\frac{\rd k}{2\pi k}\asr(k)\e^{ikt}
\nn\\&=&
\psi_R(t)+\frac{\chi}{2}\int\rd s \si_R(s)\sgn(t-s),\nn
\eea
 as well as $R\to L$.

It  is perhaps  useful to note that
 formally we can write
\bea
\psi_{\sigma R}(t)&=&Y_R\psi_R(t)Y_R^\dagger,\nn
\eea
where
\[
Y_R:=\exp\left(
\ri\chi\int\rd t\sigma_R(t)\psi_R(t)\right)
\]
Note that $Y_R$ is not a well defined operator, since $\hat\sigma_R(0){\neq}0$.

Expressed in position representation the fields are
given by
\bea
 \phi( g_R, g_L)&=&\int\rd t
 (g_R(t)-\hg0\si_R(t))\psi_{\sigma R}(t)\nn\\
 &&+R\to L
 +\hg0 p.
\label{phigenx}
\eea
Since $\int(g_R(t)-\hg0\si_R(t))\rd t=0$ the whole expression is well
defined.

The
commutator of two fields equals
\bea
[\phi(f_1),\phi(f_2)]&=&\ri\int \rd t_1\rd t_2\rd x_1\rd x_2
f_1(t_1,x_1)f_2(t_2,x_2)\nn\\
&&\!\!\!\!\!\!\!\!\times\left(\sgn(t_1{-}t_2{+}x_1{-}x_2){+}
\sgn(t_1{-}t_2{-}x_1{+}x_2)\right)\nn
\eea
Note that the commutator of fields is causal -- it  vanishes if
the supports of $f_1$ and $f_2$ are spatially separated.

\section{The $SL(2,{\mathbb R})\times SL(2,{\mathbb R})$  covariance}

Massless fields in 1+1 dimension satisfying the constraint
(\ref{cons}) actually possess much bigger
symmetry than just the $A_+(1,{\mathbb R})\times A_+(1,{\mathbb R})$ symmetry,
 they are covariant wrt the action
of $SL(2,{\mathbb R})\times SL(2,{\mathbb R})$ (for right and left
movers).

We will restrict ourselves to the action of $SL(2,{\mathbb R})$
for, say, right movers. First we consider it on the level of test
functions.

We assume that test functions satisfy $\hat g(0)=0$ and
\beq
g(t)=O(1/t^2),\ \ \ |t|\to\infty.
\label{dad1}
\eeq
Let
\beq
 C=
\left[\begin{array}{ll} a&b \\ c&d
\end{array}\right]\in  SL(2,{\mathbb R})
\eeq
(i.e. $ad-bc=1$).
We define the  action of $C$ on $g$ by
\beq
(r_C g)(t)=(-ct+a)^{-2}g\left(\frac{dt-b}{-ct+a}\right).
\label{acti}
\eeq
Note that  (\ref{acti})  preserves (\ref{dad1}) and the scalar
product
\[
(r_Cg_1|r_Cg_2)=(g_1|g_2),
\]
and is a representation, that is $r_{C_1}r_{C_2}=r_{C_1C_2}$.

We second quantize $r_C$ by introducing the unitary operator
$R_R(C)$ on ${\mathcal H}_R$ fixed uniquely by the
conditions
\[
R_R(C)\Omega_R=\Omega_R,
\]
\beq
R_R(C)\psi_R(t)R_R(C)^\dagger=\psi_R\left(\frac{at+b}{ct+d}\right).
\label{sl2rac}\eeq
Note that
\bea&& R_R(C)\left(\int \rd t g_R(t)\psi_R(t)\right)R_R(C)^\dagger
\nn\\&& \ \ \ \ =\int \rd t( r_Cg_R)(t)\psi_R(t).\eea
$C\mapsto R_R(C)$ is a  representation in ${\mathcal H}_R$.
Thus the operators $R_R(C)$ act naturally on fields satisfying
(\ref{cons}) (and hence also $\hat g_R(0)=0$). 

The fields without the constraint (\ref{cons}) are not covariant
with respect to  $SL(2,{\mathbb R})\times SL(2,{\mathbb R})$, since this
symmetry fails even at the classical level.
What remains is the $A_+(1,{\mathbb R})\times A_+(1,{\mathbb R})$ symmetry
described in (\ref{koko}). Note that 
$A_+(1,{\mathbb R})$ 
can be viewed as a subgroup of $SL(2,{\mathbb R})$:
\bea A_+(1,{\mathbb R})\ni(a,b)&\mapsto&
 C\nn\\&=&
\left[\begin{array}{ll} a^{1/2}&ba^{-1/2} \\ 0&a^{-1/2}
\end{array}\right]\in SL(2,{\mathbb R}).\label{idi}\eea
Clearly,  on the restricted Hilbert
space,   under the
identification (\ref{idi}), 
$R_R(a,b)$ coincides with $R_R(C)$.


\section{Normal ordering}

In the theory without the compensating sector the normal ordering can be
introduced in a standard way. In particular we have
\beq
{:}\,\e^{\ri\phi(g_R,g_L)}{:}\ =\
\e^{\frac12(g_R|g_R)+\frac12(g_L|g_L)}
\e^{\ri\phi(g_R,g_L)}
\label{normord}
\eeq

If the compensating sector is present then the theory does not act in the
Fock space any longer and  we do not have an invariant particle number
operator.
It is however possible (and useful) to introduce the notion of the normal
ordering. For Weyl operators it is by definition given by
(\ref{normord}). For an arbitrary operator, we first decompose it in
terms of Weyl operators, and then we apply
(\ref{normord}).
Note that our definition has an invariant meaning wrt the change of the
compensating function: in the notation of (\ref{upoleu}) we have
$$
U:\e^{\i\phi_\sigma(g_R,g_L)}:U^{-1}
=:\e^{\i\phi_{\tilde\sigma}(g_R,g_L)}:.
$$

Normal ordering is  Poincar\'e invariant but
suffers anomalies under remaining 
 $A_+(1,{\mathbb R})\times A_+(1,{\mathbb R})$
 transformations (because the
prefactor on the rhs of (\ref{normord}) is invariant only under the
Poincar\'e group). If the constraint
(\ref{cons}) is satisfied, then normal ordering is  $SL(2,{\mathbb R})\times
SL(2,{\mathbb R})$ covariant.

\section{Classical fields}

In order to better understand massless quantum fields in 1+1 dimension it is
useful to study the underlying  classical system, that is the wave equation
in 1+1 dimension (\ref{eqofm}).

From the general representation of any classical solution
\beq
\phi(t,x)=\phi_R(t-x)+\phi_L(t+x)
\eeq
we get (in notation where $f(\pm\infty)$ stands for
$\lim_{t\to\pm\infty} f(t)$)
\bea
&&\phi(t,\infty)+\phi(t,-\infty)=\nn\\
&&\phi_R(-\infty)+\phi_L(\infty)
+\phi_R(\infty)+\phi_L(-\infty)\nn\\
&&=\phi(\infty,x)+\phi(-\infty,x)
\eea

It will be convenient to denote by the space of Schwartz functions
on ${\mathbb R}$ by ${\mathcal S}$
 and by $\partial_0^{-1}{\mathcal S}$ the space of
functions whose derivatives belong to  ${\mathcal S}$ and
satisfy
 the condition $f(\infty)=-f(-\infty)$. 

We are interested only in those solutions that restricted to lines
of constant time and lines of constant position belong to
$\partial_0^{-1}{\mathcal S}$ (we will denote them  as $\cF_{11}$).
Neglecting a possible global constant shift we therefore assume that they
satisfy
\beq
\phi(t,\infty)+\phi(t,-\infty)=\phi(\infty,x)+\phi(-\infty,x)=0
\eeq
$\cF_{11}$ is characterized by two numbers
\bea\lim_{t\to\infty}\phi(t,x)&=-\lim\limits_{t\to-\infty}\phi(t,x)&=:c_0,\nn\\
\lim_{x\to\infty}\phi(t,x)&=-\lim\limits_{x\to-\infty}\phi(t,x)&=:c_1.\nn
\eea
It is natural to distinguish the following subclasses of solutions to 
(\ref{eqofm}):
\begin{itemize}
\item
$\cF_{00}$ -- solutions that restricted to lines of constant time
and to lines of constant position belong to ${\mathcal S}$ i.e.
$c_0=c_1=0$.
\item
$\cF_{10}$ -- solutions that restricted to lines of constant time
belong to  ${\mathcal S}$ and restricted to lines of constant
position belong to $\partial_0^{-1}{\mathcal S}$ i.e. $c_1=0$.
\item
$\cF_{01}$ -- solutions that restricted to lines of constant
position belong to  ${\mathcal S}$ and restricted to lines of
constant time belong to $\partial_0^{-1}{\mathcal S}$ i.e. $c_0=0$.
\end{itemize}

There are several useful ways to parametrize elements of $\cF_{11}$.
\begin{enumerate}\item {\bf Initial conditions at $t=0$:}
\begin{equation}\begin{array}{rll} f_0(x)&=&\phi(0,x),\\
f_1(x)&:=&\partial_t\phi(0,x).\end{array}\label{ini}
\end{equation}
Here,  $f_0\in\partial_0^{-1}{\mathcal S}$, $f_1\in{\mathcal S}$. Note 
that
\bea
c_0=\frac12\int f_1(x)\rd x,&\ \ \ \ &c_1=f_0(\infty).\nn
\eea
\item
{\bf Derivatives of right/left movers:}
\bea g_R(t)&:=&-\frac12 f_0'(-t)+\frac12 f_1(-t),\nn\\
g_L(t)&:=&\frac12 f_0'(t)+\frac12 f_1(t).\nn\eea
Note that $g_R,g_L\in{\mathcal  S}$ and  they
satisfy
\bea
\int g_R(t)\rd t=c_0-c_1,&
\ \ &\int g_L(t)\rd t=c_0+c_1.
\label{const}
\eea
\item {\bf Right/left movers:}
\bea
\phi_R(t)
&=&\frac12\int g_R(t-u)\sgn(u)\rd u\nn\\
\phi_L(t)&=&\frac12\int g_L(t-u)\sgn(u)\rd u.\label{solu}
\eea
Note that  $\phi_R,\phi_L\in\partial_0^{-1}{\mathcal  S}$ and they
satisfy
\bea
\phi_R(\infty)&=-\phi_R(-\infty)&=\frac12(c_0-c_1),\nn\\
\phi_L(\infty)&=-\phi_L(-\infty)&=\frac12(c_0+c_1).
\label{const1}
\eea
\end{enumerate}
We can go back from $(g_R,g_L)$ to $(f_0,f_1)$ by
\bea
f_0(x)&=&\frac12\int g_R(s-x)\sgn(-s)\rd s\nn\\&&+
\frac12\int g_R(s+x)\sgn(-s)\rd s,\nn\\
f_1(x)&=&g_R(-x)+g_L(x).\nn
\eea
We can go back from $(\phi_R,\phi_L)$  to $(g_R,g_L)$
by
\[g_R=\phi_R',\ \ \ g_L=\phi_L'.\]

The
unique solution of  (\ref{eqofm}) with  the initial conditions
(\ref{ini}) equals
\[ \phi(t,x)=\phi_R(t-x)+\phi_L(t+x).\]
It will be sometimes denoted by $\phi(g_R,g_L)$.

In the literature, one can find all three parametrizations of
solutions of the wave equation. In particular, note that 3. is
especially useful in the case of $\cF_{00}$, since then
$\phi_R,\phi_L\in{\mathcal S}$.

Note that in our paper we use 2. as the standard parametrization
of solutions of the wave equation. We are interested primarily in
the space $\cF_{10}$.  Note that $\cF_{10}$ are the solutions to
the wave equation with  $f_0,f_1\in{\cal S}$. Equivalently, for
$\cF_{10}$, the functions $g_R$, $g_L$ satisfy
\beq
\int g_R(t)\rd t=\int g_L(t)\rd t.
\label{fad}
\eeq

We equip the space  $\cF_{11}$ with the Poisson bracket, which we write for
all three parametrizations:
\bea
&&\{\phi(g_{R1},g_{L1}),\phi(g_{R2},g_{L2})\}\nn\\
&=&
\int f_{01}(x) f_{12}(x)\rd x-
\int f_{02}(x) f_{11}(x)\rd x,\label{piy}\\
&=&\int g_{R1}(t)\sgn(s-t) g_{R1}(s)\rd t\rd s\nn\\
&&+\int g_{L1}(t)\sgn(s-t) g_{L1}(s)\rd t\rd s\nn\\
&=&\Im(g_{R1}|g_{R2})+\Im(g_{L1}|g_{L2})\label{gdj}\\
&=&\frac12\int \partial_t\phi_{R1}(t)\phi_{R2}(t)\rd t\nn\\
&&+\frac12\int \partial_t\phi_{L1}(t)\phi_{L2}(t)\rd t.\label{gdj1}
\eea
Above,  $(f_{0i},f_{1i})$ and $(\phi_{Ri},\phi_{Li})$
correspond to $(g_{Ri},g_{Li})$.
The formula in (\ref{piy}) is the usual Poisson bracket for the space of
solutions of relativistic 2nd order equations (both wave and Klein-Gordon
equations). (\ref{gdj}) we have already seen in (\ref{phig}).

The Poisson bracket in $\cF_{11}$ is invariant wrt to the conformal group
fixing the infinities preserving separately the orientation of right and 
left movers, 
that is $\Diff_+({\mathbb R})\times\Diff_+({\mathbb R})$. In the case of
$\cF_{00}$  we can extend this action to the full orientation 
preserving conformal group, that is
$\Diff_+(S^1)\times\Diff_+(S^1)$, where we identify ${\mathbb R}$ together 
with the point
at infinity with the unit circle.


\section{Algebraic approach}

Among mathematical physicists, it is popular to use the formalism of
$C^*$-algebras to describe quantum systems. A description of
massless fields in 1+1 dimension within this formalism is sketched in this
section.

To quantize the space $\cF_{11}$,  we consider formal expressions
\bea \e^{\ri\phi(g_R,g_L)}\label{wey}
\eea
 equipped with the  relations
\bea
 \e^{\ri\phi( g_{R1}, g_{L1})}\e^{\ri\phi( g_{R2}, g_{L2})}&=&
\e^{\ri\Im(g_{R1}|g_{R2})
+\ri\Im(g_{L1}|g_{L2})}\nn
\\&&\times \e^{\ri\phi( g_{R1}+g_{R2}, g_{L1}+g_{L2})}
;\nn\\
 (\e^{\ri\phi(g_R,g_L)})^\dagger&=& \e^{\ri\phi(-g_R,-g_L)}.\nn
\eea
Linear combinations of (\ref{wey}) form a $*$-algebra, which we will denote
$\Weyl(\cF_{11})$.  (If we want, we can
take its completion in the natural norm and obtain a $C^*$-algebra).

Note that the group  $\Diff_+({\mathbb R})\times\Diff_+({\mathbb R})$
 act on $\Weyl(\cF_{11})$ by $*$-automorphisms.
In other words,
we have two actions
\bea
\Diff_+({\mathbb R})\ni F\mapsto\alpha_R(F)\in  
\Aut(\Weyl(\cF_{11})),\nn\\
\Diff_+({\mathbb R})\ni F\mapsto\alpha_L(F)\in  \Aut(\Weyl(\cF_{11})),\nn
\eea
commuting with one another given by
\begin{equation}\begin{array}{l}
\alpha_R(F)\left( \e^{\ri\phi(g_R,g_L)}\right)
= \e^{\ri\phi(r_Fg_R,g_L)},\\
\alpha_L(F)\left( \e^{\ri\phi(g_R,g_L)}\right)
= \e^{\ri\phi(g_R,r_Fg_L)}.\end{array}\label{auto}\end{equation}
Above,
$\Aut(\Weyl(\cF_{11}))$ denotes the group of $*$-automorphisms of the algebra
$\Weyl(\cF)$) and $r_Fg(t):=\frac1{F'(t)}g(F^{-1}(t))$.

Similarly $\Diff_+(S^1)\times\Diff_+(S^1)$  acts on $\Weyl(\cF_{00})$ by
 $*$-automorphisms.

The state $\omega$ given by  (\ref{qwq}) is
invariant wrt $ A_+(1,{\mathbb R})\times  A_+(1,{\mathbb R})$
 on $\Weyl(\cF_{11})$ and
wrt $SL(2,{\mathbb R})\times SL(2,{\mathbb R})$ on $\Weyl(\cF_{00})$.

In our paper we restricted
 ourselves to $\Weyl(\cF_{10})$.

The constructions presented in this paper give
 representations of  $\Weyl(\cF_{10})$
in a Hilbert space $\cH$ and
two  commuting with one another
strongly continuous unitary representations
\bea
 A_+(1,{\mathbb R})\ni (a,b)\mapsto R_R(a,b)\in U(\cH),\nn\\
 A_+(1,{\mathbb R})\ni (a,b)\mapsto R_L(a,b)\in U(\cH).\nn\eea
implementing the automorphisms
(\ref{auto}):
\bea
\alpha_R(a,b)(A)=R_R(a,b)AR_R(a,b)^\dagger,\\
\alpha_L(a,b)(A)=R_L(a,b)AR_L(a,b)^\dagger.\eea
In the case of the algebra $\Weyl(\cF_{00})$ the same is true for $
 SL(2,{\mathbb R})$.

In section
\ref{sec.hil} we described two  representations that satisfy the above
mentioned conditions.
 The first, call it $\pi_{\rm I}$,  represents
 $\Weyl(\cF_{10})$
 in a separable Hilbert space. Its drawback is the
absence of a vacuum vector -- a Poincar\'{e}  invariant vector.
The second
call it $\pi_{\rm II}$,  represents
 $\Weyl(\cF_{10})$
 in a non-separable Hilbert space. It has an invariant vector $|\Omega\rangle$.

We can perform the GNS construction with $\omega$. As a result we obtain the
representation $\pi_{\rm II}$ together with the cyclic invariant vector
$\Omega$. The description of this construction for massless fields in 1+1
dimension
can be found in \cite{AMS1}, Sect. III D, and \cite{AMS2} Sect. 4.
 Note, however, that we have not seen the
representation $\pi_{\rm I}$ in the literature, even though one can argue that it
is in some ways superior to $\pi_{\rm II}$.

Let us make a remark concerning the role played by the functions
$(\sigma_R,\sigma_L)$.
We note that $\cF_{00}$ is a subspace of $\cF_{10}$ of
codimension $1$. Fixing $(\sigma_R,\sigma_L)$ satisfying (\ref{sidef}) allows
us to identify $\cF_{10}$ with $\cF_{00}\oplus{\mathbb R}$.
Thus any $(g_R,g_L)$ satisfying
(\ref{fad}) is
decomposed into the direct sum of
$(g_R-\hat g(0)\sigma_R,g_L-\hat g(0)\sigma_L)$ and $\hat
g(0)(\sigma_R,\sigma_L)$.

Of course, similar constructions can be performed for the algebra
$\Weyl(\cF_{11})$ or $\Weyl(\cF_{01})$. In the literature, algebras of
observables based on $\cF_{01}$
appear in the context of ``Doplicher-Haag-Roberts charged sectors''
in \cite{StW,Bu,BV}.

\section{Vertex operators}

Finally, let us make some comments about the so-called vertex
operators, often used in string theory \cite{GSW}. Let $\delta_y$
denote the delta function at $y\in{\mathbb R}$.

Let $t_{R1},\dots,t_{Rn}\in{\mathbb R}$ correspond to insertions
for right movers and  $t_{L1},\dots,t_{Lm}\in{\mathbb R}$
correspond to insertions for left movers. Suppose that the complex
numbers $\beta_{R1},\dots,\beta_{Rn}$, and
$\beta_{L1},\dots,\beta_{Lm}$ denote the corresponding insertion
amplitudes and satisfy
\[
\sum\beta_{Ri}=\sum\beta_{Lj}.
\]
Then the corresponding  vertex operator is formally defined as
\bea
&V(t_{R1},\beta_{R1};\dots;t_{Rn},\beta_{Rn};
t_{L1},\beta_{L1};\dots;t_{Lm},\beta_{Lm})&\nn\\ &:=
\:
  {:}\exp\left(\ri\phi(g_R,g_L)\right){:},&
\label{ver}
\eea
where
\bea
g_R=\beta_{R1}\delta_{t_{R1}}+\cdots+
\beta_{Rn}\delta_{t_{Rn}},&&
\label{vee1}\\
g_L=\beta_{L1}\delta_{t_{L1}}+\cdots+
\beta_{Lm}\delta_{t_{Lm}}.&&
\label{vee2}
\eea

Strictly speaking, the rhs of (\ref{ver}) does not make sense as
an operator in the Hilbert space. In fact, in order that
$\e^{\ri\phi(g_R,g_L)}$ be a well defined operator,  we need
that
\bea
&\int\frac{\rd k}{2\pi k}
 \left|\hgR(k)-\hg0\hsi_R(k)\right|^2&\nn\\ & +
\int\frac{\rd k}{2\pi k}
 \left|\hgL(k)-\hg0\hsi_L(k)\right|^2&<\infty.
 \label{ver1}
 \eea
This is not satisfied if $g_R$ or $g_L$ are as in (\ref{vee1})
and (\ref{vee2}).

Nevertheless, proceeding formally,  we  can  deduce various
identities. For instance, we have the Poincar\'{e}  covariance:
\bea
&&R_R(a,b_R)R_L(a^{-1},b_L)\nn\\
&\times&V(t_{R1},\beta_{R1};\dots;t_{Rn},\beta_{Rn};
t_{L1},\beta_{L1};\dots)
\nn\\
&\times& R_L^\dagger(a^{-1},b_L)R_R^\dagger(a,b_R)\nn\\
&=&
V({a t_{R1}+b_L},\beta_{R1};\dots;{a^{-1} t_{Rn}+b_R},\beta_{Rn};\nn\\
&&a^{-1}t_{L1}+b_L,\beta_{L1};\dots).\nn
\eea
If in addition $\sum\beta_{Ri}=0$, 
then a similar identity is true for 
$SL(2,{\mathbb R})\times SL(2,{\mathbb R})$.

Clearly, we have
\bea
&&\omega\left(
V(t_{R1},\beta_{R1};\dots;t_{Rn},\beta_{Rn};
t_{L1},\beta_{L1};\dots;t_{Lm},\beta_{Lm})\right)
\nn\\[3mm]
&=&
\left\{\begin{array}{ll}1
,&\sum \beta_{Ri}=0;\\[3mm]
0,&\sum \beta_{Ri}\neq0.
\end{array}\right..
\label{veri2}
\eea

The following identities are often  used in string theory for the
calculation of on--shell amplitudes. Suppose that
$t_{R1},\dots,t_{Rn}$ are distinct, and the same is true for
$t_{L1},\dots,t_{Ln}$. Then, using (\ref{wx}), we obtain
\bea
&&V(t_{R1},\beta_{R1};t_{L1},\beta_{L1})\cdots
V(t_{Rn},\beta_{Rn};t_{Ln},\beta_{Ln})
\nn\\[2mm]
&&=
\e^{\left(\sum\limits_{i<j}{\cal W}(t_{Ri}-t_{Rj}))\beta_{Ri}\beta_{Rj}+
{\cal W}(t_{Li}-t_{Lj})\beta_{Li}\beta_{Lj}\right)}
\nn\\[2mm]
&&\times V(t_{R1},\beta_{R1};\dots;t_{Rn},\beta_{Rn};
t_{L1},\beta_{L1};\dots;t_{Ln},\beta_{Ln})\nn\\[2mm]
&&=
\prod\limits_{i<j}
\Big(\frac{t_{Ri}-t_{Rj}}
{\ri\mu\e^{\gamma_E}}\Big)^{-\beta_{Ri}\beta_{Rj}/2\pi}
\Big(\frac{t_{Li}-t_{Lj}}
{\ri\mu\e^{\gamma_E}}\Big)^{-\beta_{Li}\beta_{Lj}/2\pi}
\nn\\[2mm]
&&\times V(t_{R1},\beta_{R1};\dots;t_{Rn},\beta_{Rn};
t_{L1},\beta_{L1};\dots;t_{Ln},\beta_{Ln}).
\nn
\eea

\section{Fermions}

Massless fermions in 1+1 dimension do not pose such problems as bosons.
The fields are spinors, they will be written as
$\left[\begin{array}{c}\lambda_R(t,x)\\ \lambda_L(t,x)\end{array}\right]$.
 They satisfy the Dirac equation
\[\left[\begin{array}{cc}
\pa_t-\pa_x&0\\
0&\pa_t+\pa_x\end{array}\right]
\left[\begin{array}{c}\lambda_R(t,x)\\ \lambda_L(t,x)\end{array}\right]=0.\]

We will also use the fields smeared with real functions~$f$, where the
condition (\ref{cons}) is not needed any more:
\[
\left[\begin{array}{c}\lambda_R(f)\\ \lambda_L(f)\end{array}\right]
=\int\left[\begin{array}{c}\lambda_R(t,x)\\ \lambda_L(t,x)\end{array}\right]
f(t,x)\rd t\rd x.
\]
Because of the Dirac equation, they can be written as
\[\lambda_R(f)=\lambda_R(g_R),\ \ \lambda_L(f)=\lambda_L(g_L),\]
where $g_R$ and $g_L$ where introduced when we discussed bosons.

For $k>0$, we introduce fermionic operators (for right and left sectors)
$b_R(k)$ and $b_L(k)$ satisfying the anticommutation relations
\bea
\{b_R(k),b_R^\dagger(k')\}&=&2\pi\de(k-k'),\nn\\
\{b_L(k),b_L^\dagger(k')\}&=&2\pi\de(k-k'),
\eea
with all other anticommutators vanishing.
Now
\bea
\la_R(g_R)
&=&\int\frac{\rd k}{2\pi}\left(g_R^\ast(k)
b_R(k)+g_R(k)b_R^\dagger(k)\right)
\nn\\
\la_L(g_L)
&=&\int\frac{\rd k}{2\pi}\left(g_L^\ast(k)
b_L(k)+g_L(k)b_L^\dagger(k)\right)
\eea
The anticommutation relations for the smeared fields read
\bea
\{\lambda_R^\dagger(g_{R1}),\lambda_R(g_{R2})\}&=&\int g^*_{R1}(t)
g_{R2}(t)\rd
t\nn\\
&=&\int \frac{\rd k}{2\pi}\hat g^*_{R1}(k)\hat g_{R2}(k)
\label{ggferm}
\eea
and similarly for the left sector.
Note the difference of the fermionic scalar product (\ref{ggferm}) and the
bosonic one $(\cdot|\cdot)$.

In terms of space-time
smearing functions these anticommutation relations read
\bea
\{\lambda_R^\dagger( f_{1}),\lambda_R(f_{R})\}=2\int\rd t\rd x\delta(t+x)
f_{1}(t,x)f_{2}(t,x),\nn\\
\{\lambda_L^\dagger (f_{1}),\lambda_R(f_{L})\}
=2\int\rd t\rd x\delta(t-x)
f_{1}(t,x)f_{2}(t,x).\nn\eea

Fermionic fields are covariant with respect to the group
$ A_+(1,{\mathbb R})\times  A_+(1,{\mathbb R})$. We will restrict ourselves to
discussing the covariance for say, right movers. The right
Hamiltonian and the right dilation generator are
\bea
H_R^\rf&=&\int\frac{\rd k}{2\pi}k
b_R^\dagger(k)b_R(k)
\nn\\
D_R^\rf&=&\frac{\ri}{2}\int\frac{\rd k}{2\pi}
\left(b_R^\dagger(k)k\pa_kb_R(k)
-
\bigl(k\pa_kb_R^\dagger(k)\bigr)b_R(k)
\right).\nn
\eea
We have the usual commutation relations for $H_R^\rf$ and $D_R^\rf
$ and their action on the fields is anomaly-free:
\bea
 \left[H_R^\rf,\lambda_R( g_R)\right]&=&-\ri\lambda
(\pa_tg_R),\nn\\
  \left[ D_R^\rf,\lambda_R(g_R)\right]&=&\ri\lambda\bigl(
(t\pa_t +1/2) g_R\bigr).
 \nn\eea

We have also the covariance with respect to the  conformal group
$SL(2,{\mathbb R})\times SL(2,{\mathbb R})$.
 We need to assume
that test functions satisfy
\beq
g(t)=O(1/t),\ \ \ |t|\to\infty.
\label{dadfe}
\eeq
We define the  action of
\beq
 C=
\left[\begin{array}{ll} a&b \\ c&d
\end{array}\right]\in  SL(2,{\mathbb R})
\eeq
on $g$ by
\beq
(r_C^\rf
 g)(t)=(-ct+a)^{-1}g\left(\frac{dt-b}{-ct+a}\right).
\label{actife}
\eeq
Note that  (\ref{actife}) has a different
power than (\ref{acti}). It  is a unitary representation for
the scalar
product $(\cdot|\cdot)_\rf$.

We second quantize $r_C^\rf$ on the fermionic Fock space by
introducing the unitary operator $R_R^\rf(C)$ fixed
uniquely by the conditions
\[
R_R^\rf(C)\Omega_R=\Omega_R,
\]
\beq
R_R^\rf(C)\lambda_R(t)R_R^\rf(C)^\dagger
=(ct+d)^{-1}\lambda_R\left(\frac{at+b}{ct+d}\right).
\label{sl2racfe}
\eeq
Note that $C\mapsto R_R^\rf(C)$ is a  unitary representation
 and it acts naturally on fields:
\bea
&& R_R^\rf(C)\lambda_R(g_R)
R_R^\rf(C)^\dagger
=\lambda_R(r_C^\rf g_R),
\eea

\section{Supersymmetry}

In this section we consider both bosons and fermions. Thus our
Hilbert space is the tensor product of the bosonic and fermionic
part. We assume that the bosonic and fermionic operators commute
with one another. Clearly, our theory is  $A_+(1,{\mathbb R})\times
 A_+(1,{\mathbb R})$ covariant. In
fact, the  right Hamiltonian and the generator of dilations for
the combined theory are equal to $H_R+H_R^\rf$ and $D_R+D_R^\rf$.

In the case of the theory with the constraint (\ref{cons}),
we have also the $SL(2,{\mathbb R})\times
SL(2,{\mathbb R})$. covariance.

On top of that, the combined theory is supersymmetric.
The supersymmetry generators $Q_R$, $Q_L$ are defined as
\bea
Q_R&=&\int\frac{\rd k}{2\pi}\left(\asr^\dagger(k)
b_R(k)+\asr(k)b_R^\dagger(k)\right)
\nn\\
Q_L&=&\int\frac{\rd k}{2\pi}\left(\asl^\dagger(k)
b_L(k)+\asl(k)b_L^\dagger(k)\right)
\eea

They satisfy the basic supersymmetry algebra relations without the
central charge
\bea
\{Q_R,Q_R\}&=&2( H_R+H_R^\rf),\nn\\
\{Q_L,Q_L\}&=&2( H_L+H_L^\rf),\nn\\
\{Q_R,Q_L\}&=&0.
\eea
The action of the supersymmetric charge transforms bosons into fermions
 and vice versa:
\bea
\left[Q_R,\phi(g_R,g_L)\right]&=&\lambda_R(g_R),\nn
\\
\left[Q_L,\phi(g_R,g_L)\right]&=&
\lambda_L(g_L),
\nn\\
\left[Q_R,\lambda_R(g_R)\right]
&=&\phi(\pa_t g_R,0),\nn\\
\left[Q_L,\lambda_L(g_L)\right]
&=&\phi(0,\pa_t g_L).\nn
\eea

The pair of operators
$\left[\begin{array}{c}Q_R\\ Q_L\end{array}\right]$
behaves like a spinor under the Poincar\'e{} group. Even more is true: we
have the covariance under the group $ A_+(1,{\mathbb R})\times
 A_+(1,{\mathbb R})$, which
for the  right movers can be expressed in terms of the following
commutation relations:
\bea
 [H_R,Q_R]&=&0,\nn \\  \   [ D_R,Q_R]&=&-\frac{\ri}{2}Q_R.
\nn\eea

\medskip

\noindent {\small{\bf Acknowledgement}
J.D. was partly supported by
the European Postdoctoral Training Program HPRN-CT-2002-0277, the Polish
KBN grant  SPUB127 and 2 P03A 027 25. K.A.M. was
partially supported by the Polish KBN grant 2P03B 001 25 and the
European Programme HPRN--CT--2000--00152.

J.D.  would like to thank S.~DeBi\`evre, C.~G\'erard
 and C.~J\"akel for useful discussions. 

\noindent
{\it E-mail addresses:}\\
\noindent
{\tt Jan.Derezinski@fuw.edu.pl},\\ {\tt
Krzysztof.Meissner@fuw.edu.pl}

\end{document}